\documentclass[%
 preprint,
nofootinbib,
 amsmath,amssymb,
floatfix,
]{revtex4-1}

\usepackage{graphicx}
\usepackage{dcolumn}
\usepackage{bm}
\usepackage{braket}
\usepackage{subfig}
\usepackage[style=base]{caption}
\usepackage{hyperref}

\begin{document}

\preprint{arXiv:1611.08581}

\title{Towards a dS/MERA correspondence}

\author{Raj Sinai Kunkolienkar}
 \email{raj.kunkolienkar@gmail.com}
 
\author{Kinjal Banerjee}%
 \email{kinjalb@gmail.com}
\affiliation{%
 BITS, Pilani -- K. K. Birla Goa Campus \\
 Zuarinagar, Goa -- India 
}%

\date{\today}

\begin{abstract}
Recent advances have suggested that spacetime itself emerges from the entanglement of the quantum degrees of freedom living on the boundary. In the case of the AdS spacetimes, a particular class of tensor networks has been shown to realize the same via Multi-Scale Entanglement Renormalization Ansatz(MERA). In this paper we suggest a prescription for the dS/MERA correspondence and recover a discrete version of de Sitter Penrose diagram by using the MERA on conformal theories identified with the future/past boundaries ($\mathcal{I^\pm}$) of the de Sitter spacetime. In this case, as anticipated, time appears as the emergent direction. We comment on the possible interpretation that the de Sitter cosmological horizon entropy involves entanglement with degrees of freedom across the cosmological horizon as well as the implications of our construction for cosmology.
\end{abstract}

\pacs{Valid PACS appear here}
\maketitle


\section{Introduction}
\label{sec:intro}

Recently, an interesting correspondence was conjectured linking the Multi-Scale Entanglement Renormalization Ansatz (MERA) network to the AdS/CFT correspondence \cite{Swingle:2009bg}. Tensor networks for holography serve as good intuitive guides and make for good toy models wherein the holographic correspondence has been seen to go beyond just reproducing the bulk geometry as the network geometry. In parallel, there were ideas that put forth the view that as a general rule, spacetime itself emerges from quantum entanglement on the boundary \cite{VanRaamsdonk:2010pw}. In the same spirit, one should probably expect to find a tensor network formalism or prescription which describes a general spacetime \footnote{Not all spacetimes may allow for a holographic description -- not all quantum superpositions give a classical geometry, and the question itself is not well understood yet, but we shall go ahead with the assumption that there may exist a holographic description for the de Sitter spacetimes and try to find a prescription for it.}. To start, one should consider the spacetimes whose descriptions form the simplest solutions of the Einstein Field Equations -- spacetimes which are maximally symmetric. With such a correspondence for the anti-de Sitter (AdS) spacetime seemingly under our belt, we turn our attention towards the de Sitter (dS) spacetimes. The need to theoretically tackle the description of de Sitter spacetimes is of prime importance not only because it is a maximally symmetric spacetime but also due to its cosmological relevance. Observations have long suggested that the universe that we inhabit had not only undergone a quasi de Sitter phase in its early history \cite{Baumann:2009ds} but that it would also go on to do so at later times \cite{Riess:1998cb}. 

Historically, attempts have been made to try and postulate a de Sitter analog for the holographic duality, beginning with the hypothesized dS/CFT correspondence \cite{Strominger:2001pn}. In this note, we would like to suggest a prescription for a dS/MERA correspondence, which is strongly motivated by the original dS/CFT correspondence as well as derivative work done on it. Our work is strongly motivated by the fact that the metric on the MERA network geometry is identical in form to the one written for a particular slicing of de Sitter spacetime. Through our construction, we go on to comment on the origins of the entropy of the de Sitter cosmological horizons. Though one has not been able to find a consistent stringy realisation of holography in de Sitter, we hope that our top-down approach might help us approach the problem from a new perspective. We note that the lack of a concrete understanding of holography in de Sitter spacetimes \cite{Anninos:2012qw} inhibits us from drawing strong conclusions about our suggestions.

As a very quick review of the ingredients that go into our prescription, we introduce the de Sitter spacetime in (Sec.\ref{de Sitter}) and go on to describe the various coordinate systems that would be of use to us. We also give a brief description of the ideas pertaining to thermodynamics of the de Sitter spacetime. The section ends by introducing the idea of holography in the AdS/CFT correspondence as well and extending it to the various proposals made in the context of de Sitter holography. In (Sec.\ref{sectionMERA}), we introduce the idea of the MERA as a scheme and go on to illustrate the relevance and use of the tensor network in the context of the AdS/CFT correspondence. We follow it up with the arguments which give us the metric and the casual structure of the MERA as well as the implementation of the MERA for a thermal state. These crucial points lead us to our prescription in (Sec.\ref{sec4}) which essentially forms the bulk of the new ideas introduced in the paper and is basis for the discussion to follow in (Sec.\ref{discussion}). We close with pointing out directions for future explorations as well as the non-trivial shortcomings of our work. The suggestions on interpreting the origins of the cosmological entanglement entropy as well as realizing cosmology in tensor networks might be of relevance to ongoing efforts in the field.

\section{Explorations in de Sitter Spacetime}
\label{de Sitter}
As is common knowledge, de Sitter spacetimes form a family of maximally symmetric solutions of the vacuum Einstein Field Equations, with a positive cosmological constant ($\Lambda > 0$). In the embedding formalism, the four-dimensional de Sitter spacetimes (d$S_4$) live in a five-dimensional Minkowski spacetime ($\mathcal{M}^{1,4}$) with the metric $ds^2 = - dX_{0}^2 + dX_{i}^2$. The hyperboloids which give us the de Sitter spacetimes are given by a constraint equation : 
\begin{equation}
    -X_{0}^{2}+\sum _{{i=1}}^{4}X_{i}^{2}= L^{2}
\end{equation}
Here, the parameter $L$ gives us the characteristic de Sitter radius and is related to the cosmological constant by $\Lambda = 3/L^2$. The symmetry group of this particular submanifold is the Lorentzian $SO(1,4)$ group. The topology of the spacetime is $R^1 \times S^3$.

\subsection{Relevant Avatars}

The de Sitter geometry that we obtained above lends itself to be sliced in many different ways, each slicing corresponding to a coordinate system. Below, we shall briefly review the coordinate systems which will be of use to us. For a comprehensive overview on the slicings of de Sitter, refer to \cite{Spradlin:2001pw,Anninos:2012qw}

\paragraph{Global Geometry} We first describe a coordinate system which covers the entire dS spacetime. Given that $d\Omega_{3}^2$ is the metric on the unit 3-sphere, the induced metric can be written in the form : 
\begin{equation}
    ds^2 = -d\tau^2 + (L^2 \cosh{\frac{\tau}{L}}) d\Omega_{3}^2
\end{equation}
We note that on constant $\tau$ slices, we end up with a 3-sphere with the radius dependent on time. These spheres are seen to shrink to a non-zero size as we go from $\tau \rightarrow -\infty$ to $\tau = 0$ and then expand again as $\tau \rightarrow +\infty$.

\begin{figure}[ht]
\centering
\includegraphics[width=0.37\textwidth]{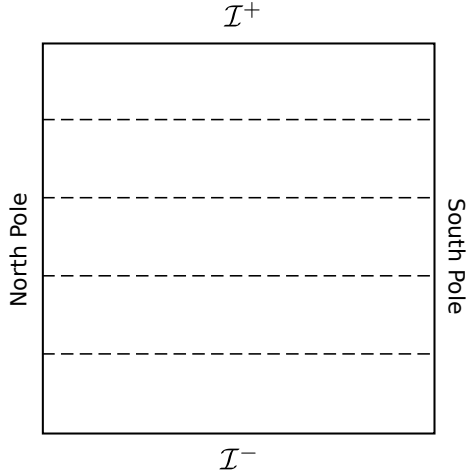}
\caption{Global coordinates on the de Sitter spacetime in $n$ dimensions ($dS_n$). The dotted lines represent surfaces of equal $T$, and this coordinate runs from $-\pi/2$ at $\mathcal{I^-}$ to $\pi/2$ at $\mathcal{I^+}$. Each point on the diagram represents a $S^{n-2}$.}
\label{globaldS}
\end{figure}

To investigate the causal structure, one needs to conformally transform the metric. To do so, we shift to new coordinates $\tan{(T/2)} = \tanh{(\tau/2L)}$. Upon doing so, we get the metric to be
\begin{equation}
    ds^2 = \frac{L^2}{\cos^2{T}} (-dT^2 + d\Omega_{3}^2)
\end{equation}
From the above equation, we can draw the Penrose diagram for the de Sitter spacetime (Figure \ref{globaldS}). Note that there exist initial $(\mathcal{I}^+)$ and final conformal boundaries $(\mathcal{I}^-)$. These surfaces are the surfaces where all null rays originate and terminate. On each constant $T$ slice there is a 3-sphere, the poles of which are denoted in the Penrose diagram as drawn for the global coordinates. 
\paragraph{Flat Slicing} There is a slicing which covers only half of the Penrose diagram, but is of significance in cosmology and for the arguments of Sec.\ref{sec4}. 
In this particular slicing, the metric is given by 
\begin{equation}
\label{dS-metric}
    ds^2 = - dt^2 + e^{2t/L} d\vec{x}^2 = \frac{L^2}{\eta^2} (-d\eta^2 + d\vec{x}^2)
\end{equation}

\begin{figure}
\centering
\subfloat[]{\includegraphics[width=.4\linewidth]{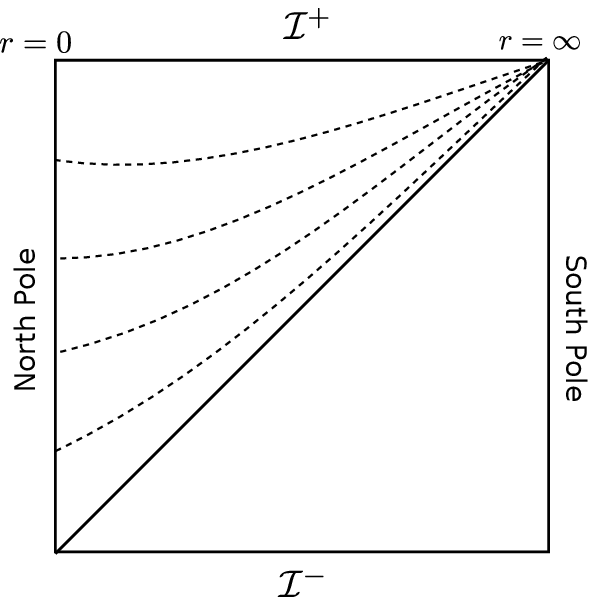}} 
\hspace{2em}
\subfloat[]{\includegraphics[width=.45\linewidth]{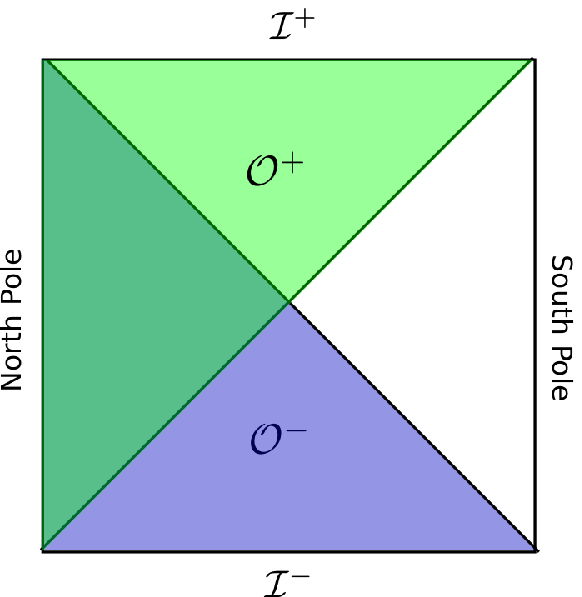}}
\caption{The figure (a) represents the planar coordinates / flat slicing for de Sitter. As can be seen, it covers only half of the spacetime. The dotted lines represent surfaces of constant cosmological time. This particular slicing is useful to describe the propagation of null rays from the North Pole. In (b), the green (blue) region can be influenced by (can influence) an observer sitting at the North Pole at $\mathcal{I}^-$ ($\mathcal{I}^+$). These regions are depicted as $\mathcal{O}^+, \mathcal{O}^-$ and the lines which bound them are known as the \textit{future event horizon} and \textit{past event horizon} respectively.} 
\label{flatfigure} 
\end{figure}

In the context of cosmology, the above metric can be recognized as that of a exponentially expanding flat FRW universe where $t$ denotes the proper time and $\eta$ gives us the conformal time. Constant $t$ or $\eta$ slices therefore give us flat spatial sections. The relation between the two coordinates is given as $e^{t/L} = -L\tau^{-1}$, where $t \in (0, \infty)$ and $\tau \in (-\infty, 0)$.
In the Penrose diagram (Figure \ref{flatfigure}), we note that the metric covers the region which is contained inside (and including) the lightcone of a point sitting on the North Pole, at $\mathcal{I}^-$. Certainly, this is the region that can be causally influenced by an observer moving along the North Pole. Note that as the observer moves from $\mathcal{I}^-$ to $\mathcal{I}^+$, the cosmological time parameter increases whereas the conformal time decreases. Alternatively, one can describe the region covered by the past light cone emanating from the observer at the North Pole, while being at $\mathcal{I}^+$. For the same form of the metric, the proper time runs in the opposite direction to that we obtain in the previous case. The boundaries of these coordinate descriptions give us the future and the past event horizons, respectively. 

\begin{figure}[ht]
\centering
\includegraphics[width=0.37\textwidth]{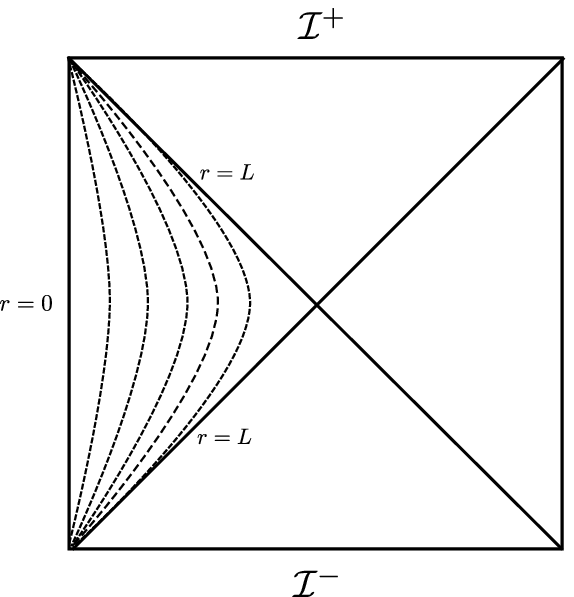}
\caption{The static patch as described in the Northern Causal Diamond. The static observer is on the North Pole ($r=0$). The dotted lines give us the surfaces of constant $r$, with the $r=L$ surface being the de Sitter cosmological horizon that we are interested in. Note how the static observer does not have access to all the data on $\mathcal{I}^+$.}
\label{staticfigure}
\end{figure}

\paragraph{Static Patch} As is evident from above the Penrose Diagram, a single observer will be unable to access the entire de Sitter spacetime. That is, he/she will not be able to have the entire spacetime in their past lightcone. This has important ramifications when one tries to deal with horizons and thermodynamics in de Sitter spacetime. The region that a single observer can influence and get influenced by ($\mathcal{O}^+ \cap \mathcal{O}^-$) is the region given by the static coordinates (Figure \ref{staticfigure}). The metric on the Northern Causal Diamond (NCD) is given as
\begin{equation}
    ds^2 = \Big( 1- \frac{r^2}{L^2} \Big) dt'^2 + \Big( 1- \frac{r^2}{L^2} \Big)^{-1} dr^2 + r^2 d\Omega_{2}^2
\end{equation}

The interesting features about this metric are the presence of a local timelike Killing vector $\partial_{t'}$ as well as the consequent vanishing of the same at $r=L$, denoting the presence of a cosmological horizon \footnote{Note that there is a difference between the cosmological horizon and the past/future event horizons, although the area of the $S^2$ surfaces which occur on any point on these null surfaces, is the same}. From the perspective of cosmology, the cosmological horizon arises because of the fact that as the universe expands more and more regions become inaccessible to the observer. Note that every observer is always surrounded by his/her horizon and unlike the case of a black hole, the observer can never reach his horizon. Note that one can similarly define an observer in the Southern Causal Diamond (SCD), with the Killing vector pointing in the direction opposite as that in the NCD. 

\subsection{Thermodynamics of de Sitter Spacetime}
Following the Bekenstein-Hawking bound on the entropy of a Schwarzschild black hole, Gibbons and Hawking calculated the temperature and the entropy of a de Sitter cosmological horizon \cite{GibbonsHawking}. Classically, this entropy is said to be a measure of the observer's lack of information of the regions which he cannot see. The temperature of the horizon is given by 
\begin{equation}
    T_\text{dS} = \frac{1}{2 \pi L} 
\end{equation}
Whereas the entropy of the cosmological horizon is given as 
\begin{equation}
    S_\text{dS} = \frac{A_\text{horizon}}{4} = \pi L^2
\end{equation}
An application of Bousso's spacelike projection theorem in the de Sitter spacetime \cite{Bousso:1999xy,Bousso:1999cb} leads us to conclude that the entropy of the past event horizon places an upper bound on the number of degrees of freedom of the volume enclosed inside the past event horizon ($N_\text{DOF} \leq \pi L^2$). The theorem relies on the fact that information will pass through the lightlike surface generated from (and bounded by) the $S^2$ surfaces on the horizon. However, the spacelike orbits of the Killing vector in the region beyond the static patch and bounded by the past event horizon do not facilitate this. Hence, this result can be translated into a bound on the degrees of freedom contained in the region bounded by the cosmological horizon -- because of the timelike nature of the Killing vector in the region. 

The microscopic origins of the de Sitter Horizon entropy have remained a mystery and over the years there have been various mechanisms that have been proposed to explain it \cite{Balasubramanian:2001rb}. One approach talks of the entropy arising from quantizing the degrees of freedom associated with the cosmological horizon \cite{Lin:1999gf}. Another competing view posits that the entropy arises due to quantum entanglement with the degrees of freedom which are inaccessible, by virtue of being hidden behind the cosmological horizon \cite{Hawking:2000da}. Motivated by recent developments \cite{VanRaamsdonk:2010pw}, \cite{Swingle:2009bg}, we will try to explore and realize the second proposal in a new way.

\subsection{Attempts at Realizing Holography}

Following the hints obtained from the Bekenstein-Hawking bound, 't Hooft and Susskind proposed the holographic principle which conjectured that data in a $D$-dimensional gravitational system could be encoded in the quantum degrees of freedom of a theory living in $D-1$-dimensions \cite{Susskind:1994vu, Bousso:2002ju}. Maldacena's AdS/CFT conjecture proved to be a concrete realisation of the same, in the context of anti de Sitter spacetimes \cite{Maldacena:2001kr}. In a stringy setting, the gravitational degrees of freedom living in the bulk are described by data CFT living on the timelike boundary. The ever expanding AdS/CFT `dictionary' talks of the duality between various quantities like the equality between the partition functions of the two theories, scattering amplitudes in AdS being described by correlation functions on the boundary CFT \cite{Witten:1998qj, Penedones:2010ue} as well as the the Ryu-Takayanagi proposal for calculating the entanglement entropy of a subregion on the boundary CFT using the area of extremal surfaces in the bulk anchored to the corresponding subregion in the bulk \cite{Ryu:2006bv}. 

Despite the successes in the AdS/CFT program, a concrete top-down holographic realisation of de Sitter spacetime has been hard to come by. As we have seen, de Sitter has two asymptotic, disconnected spacelike boundaries -- $\mathcal{I}^+$ and $\mathcal{I}^-$. This is unlike the case of AdS which has a single timelike boundary on which one defines a Lorentzian CFT (say, defined on a plane or a sphere). With an antipodal identification of points on the $S^{n-1}$ spheres living on $\mathcal{I}^+$ and $\mathcal{I}^-$, Strominger's dS/CFT conjecture proposed a link between quantum gravity in the $dS_n$ bulk and an Eucledian CFT living on a single $S^{n-1}$, by considering an analytic continuation of the AdS case \cite{Strominger:2001pn,Anninos:2011ui}. Here, correlators in the bulk whose points sit on the boundaries are linked to the CFT correlators on the sphere. It was shown that the conformal weights of the CFT \footnote{One gets $h_\pm = 1 \pm \sqrt{1-m^2L^2}$, implying that the weights are imaginary for $m^2L^2 > 1$, where $m$ is the mass of the scalar field in the bulk.} in the bulk could take on imaginary values, implying that the CFT could be non-unitary. This proposal was further made concrete in \cite{Witten:2001kn,Maldacena:2002vr,Harlow:2011ke} by stating the wavefunction of a universe which is asymptotically de-Sitter space can
be computed in terms of the partition function of the dual conformal field theory through the relation $\Psi[g] = Z[g]$. For the purpose of this work, we assume that there indeed exists a holographic description for de Sitter spacetime. 

Having seen how an observer in de Sitter does not have access to the entire spacetime, there have been debates on what would make for the correct set of observables in de Sitter. Since the observer does not have access to all the data on $\mathcal{I^+}$, some have vouched for a holographic description for the static patch, rather than the global dS/CFT correspondence that we have considered \cite{Karch:2013oqa}. The nice feature of these proposals is the ability to recover the thermal nature of the static patch. Taking inspiration from how the vacuum state in the Minkowski spacetime can be written down as a maximally entangled state constructed from excited states in the $L$ and $R$ Rindler wedges, a similar procedure was used on the de Sitter static patches\cite{Bousso:2001mw,Czech:2012be}. By tracing over one of the static patches, the thermal density matrix for the remaining patch is obtained \footnote{This analysis was done for the case of the dS$_3$/CFT$_2$ conjecture.}. Work done in \cite{Karch:2013oqa} reproduced the required entropy of the cosmological horizon for the higher spin realisation of the static patch dS - cutoff CFT conjecture, though this result comes from counting the number of degrees of freedom in the boundary theory. 

Interestingly, work done in \cite{Balasubramanian:2002zh} noted that in general, one would need two disjoint but possibly entangled CFTs with special hermiticity conditions instead of one CFT as dual to the de Sitter bulk spacetime. For the modified CFTs, denoting the Hilbert spaces of the static patches ($L$, $R$) as well as the conformal boundaries  $I^-$, $I^+$ as $\mathcal{H}_L$, $\mathcal{H}_R$, $\mathcal{H}_i$ and $\mathcal{H}_f$ respectively, it was suggested that $\mathcal{H}_L \otimes \mathcal{H}_R \equiv  \mathcal{H}_i \otimes \mathcal{H}_f$. The two bases possibly have a complicated relationship among themselves, with the de Sitter vacuum being realized as thermally entangled states in $\mathcal{H}_L \otimes \mathcal{H}_R$. The entropy of the cosmological horizon was said to arise due to a trace over the antipodal static patch in the entangled state.

Recently, work done in \cite{Verlinde:2016toy} takes on an approach on holography for de Sitter slightly different from that of the dS/CFT correspondence due to the presence of the cosmological horizon and the absence of null and spatial infinities. This view relies on the fact that the entanglement structure of de Sitter is fundamentally different from that of AdS, and the indication that microscopically de Sitter corresponds to a thermal state. However, the view adopted is not in line with the idea that the entropy arises because of entanglement with degrees of freedom $\textit{across}$ the horizon which we adopt in this paper. 

\section{The Multi-scale Renormalization Ansatz}
\label{sectionMERA}

The Multi-scale Renormalization Ansatz (MERA) belongs to a class of variational ansatz known as tensor networks. It originated as a tool in the study of quantum many body systems and condensed matter physics in order to efficiently compute the ground state of a given Hamiltonian \cite{Vidal1}. The idea underlying MERA is that of disentangling the system at various length scales as one follows coarse graining Renormalization Group(RG) flow in the system. This scheme is particularly effective for (scale invariant) critical points of the system. The MERA is endowed with a network geometry that has been crucial in realizing the violation of the area law for entanglement entropy for 1D critical systems. The matching of the network geometry with the Ryu-Takayanagi proposal led to the AdS/MERA conjecture, a discrete realisation of the Anti de Sitter spacetime \cite{Swingle:2009bg}. A rigorous introduction to the MERA can be found in \cite{hauru}.

\subsection{Constructing the MERA}

Let us review the construction of the MERA in 1D for a discrete  quantum system. A tensor with $n$ indices is represented in the network by a $n$ legged node. Contracting a particular index of a tensor with another involves joining the two legs. Hence, it can be seen that a scalar quantity will not have any free legs emanating from it. Our goal in this section would not be to demonstrate the computational efficiency of MERA, but to illustrate its construction and structure. 

Consider the general state for a quantum system with $k$ distinct sites, each site having the possibility to be in $\chi$ different states to be :
\begin{equation}
    \ket{\Psi} = \sum_{i_1, i_2 \ldots i_k = 1}^{\chi}     c_{i_1 i_2 \ldots i_k} \ket{i_1 i_2 \ldots i_k}
\end{equation}

\begin{figure}[ht]
\centering
\includegraphics[width=0.5\textwidth]{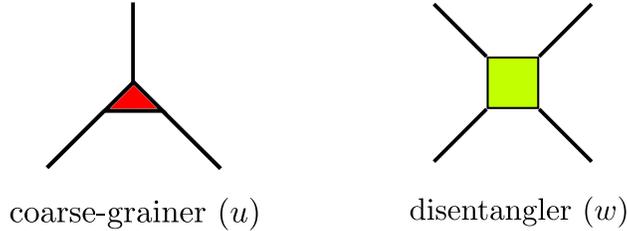}
\caption{The 3 legged coarse-grainer ($u$) takes inputs from 2 sites on a lower scale in the MERA, and gives an output onto one site which lies on a higher level in the network. The disentangler removes entanglement between two neighboring sites lying on the same level. }
\end{figure}

\begin{figure}[ht]
\centering
\includegraphics[width=0.5\textwidth]{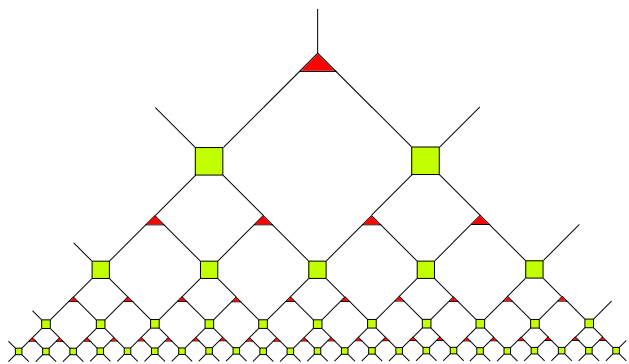}
\caption{The implementation of the MERA for a conformally invariant system. We can seen the alternating layers of disentangling and coarse-graining applications. For the network shown, we see that the number of sites after each such set of operations is halved. The MERA is seen to exhibit a fractal-like geometry with infinite levels due to the conformal nature of the system itself.}
\label{MERA}
\end{figure}

The MERA network consists of alternating layers of two distinct (unitary) tensors -- the \textit{coarse-grainer} and the \textit{disentangler}. The $n$ legged coarse-grainer ($u$) in general, has $n-1$ legs acting as the input and $1$ leg acting as the output. In the MERA picture, such a tensor is designed to effectively (exponentially) reduce the dimensions of the Hilbert space as one goes from one level to the next higher level. This acts out a scheme of tree renormalization on the quantum state. Due to the entanglement that exists at short length accumulating under this scheme, coarse-graining alone is not enough for one to get a good effective description of the system through a RG flow. To remedy this, one introduces disentanglers ($w$) which usually have 2 input legs as well as 2 output legs. The role of the disentangler is to remove the entanglement between the sites which lie at the interface of two coarse-graining blocks. In general, one ends the network from the top when one can write the coarse grained wavefunction in terms of products of pure states. For the conformally invariant case that we are interested in, the tensor network has an infinite depth.  A network implementation of such a prescription is shown in Figure (\ref{MERA}).

The relevance of the MERA to holography was explored in the context of the AdS/CFT correspondence by Brian Swingle in \cite{Swingle:2009bg, Swingle:2012wq}. This connection was facilitated by the prescription to calculate the entanglement entropy of a subregion of a 1D critical system in the MERA. To do so, one needs to trace a curve that is anchored on the boundary through the network in a way that it cuts the minimal number of disentangler links. The number of links broken goes to count the entanglement that has been discarded while setting up the density matrix for the subsystem. For a subregion of scale $l$, the entanglement entropy goes as $S_A \sim \log{l}$. Along with the idea of the extremal curve in the network geometry, this reproduces the Ryu-Takayanagi result. In higher dimensional AdS/CFT, let $\gamma_A$ give us the extremal surface anchored to $\partial A$. In the higher dimensional MERA for a critical system, let the entanglement links cut by a surface $\Omega_A$ anchored to $\partial A$ be counted by ${\partial \Omega_A}$. The entanglement entropy of the subregion in the AdS/CFT case goes as : 
\begin{equation}
    S_A = \frac{\text{Area}(\gamma_A)}{4}
\end{equation}
This result holds true in the MERA network geometry as well, since $\text{Area}(\gamma_A) \sim \partial \Omega_A$, upto a constant. Hence, one thinks of the AdS/MERA correspondence as an explicit realisation of holography given by tensor networks. However, note that Swingle's correspondence only describes constant time slices of the static AdS spacetime and not the whole of the AdS spacetime. Recent work has suggested that such tensor network correspondences should not be read off as giving us a direct construction of the bulk geometry but as being representative of structures in the `kinematic space' of the CFT \cite{Czech:2015kbp}. However, given that the work is done for the case of Lorentzian CFTs, we wouldn't wish to extrapolate the same as being true in the case of de Sitter holography as the situation with respect to the conformal boundaries and extremal curves is much more complicated. If anything, the tensor network construction that we plan to suggest could serve as a guide for similar `kinematic space' structures in the case of de Sitter holography. 

\subsection{Causal Structure of the MERA}

Due to the structure of the MERA, it is possible to define a notion of causality on the network. As first explored in \cite{Beny:2011vh}, one obtains the metric of a hyperbolic space. This was seen as a check for the above noted AdS/MERA correspondence. Since our proposal relies heavily on the causal structure of the MERA, we shall briefly review the elegant proof given below \cite{Vidal:talk}. 
\begin{figure}[ht]
\centering
\includegraphics[width=0.5\textwidth]{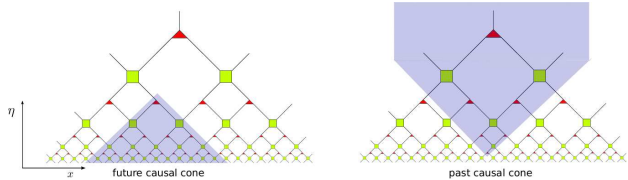}
\caption{In the right image, we see the depiction of the future causal cone for a site defined in the MERA geometry. In the left image, the past causal cone of a site in the network geometry is illustrated. The past/future conventions have been chosen with a foresight, such that they obey the $t\rightarrow\infty$ condition on $\mathcal{I}^+$ (which will be seen to coincide with the UV degrees of freedom of the system).}
\label{causal}
\end{figure}

It can be seen that one can define the notion of a `past' and `future' causal cone in the MERA. These rely on the fact that a particular tensor in the network can `get influenced' and `influence' other tensors -- mediated by the contractions that correspond to inputs and outputs (Figure \ref{causal}). Hence, we can infer that the network geometry will be endowed with a Lorentzian metric structure. For the case of the MERA for a 1D system, we shall make use of two coordinates systems -- the cosmological coordinates $(\eta, x)$ and the causal cone coordinates $(v^+, v^-)$. For reasons that will be clear soon, we shall call $\eta$ the \textit{conformal time} and $x$ as the \textit{comoving distance}. The two coordinate systems are related as : $v^\pm \equiv \eta \pm x$. 

The measure of distance that we shall adopt in the network would be such that each disentangler is the same distance away from its nearest disentanglers. To satisfy this, the most general metric for a tensor network that is homogeneous at each layer is given by :
\begin{equation}
    ds^2 = f_{\eta \eta}(\eta, x) d\eta^2 + 2 f_{\eta x} (\eta,x) d\eta dx + f_{x x}(\eta, x) dx^2
\end{equation}

Changing to the causal cone coordinates and taking cognizance of the fact that the metric should have a Lorentzian structure we obtain the metric wherein due to the null propagation along $v^+$ and $v^-$, only cross terms survive : 
\begin{equation}
    ds^2 = f(\eta, x)( dv^+ dv^-)
\end{equation}
Changing back to the cosmological coordinates, we get the metric as $ds^2 = f(\eta, x)( d\eta^2 + dx^2)$. We note that for the function $f(\eta, x)$ to agree with the homogeneity condition, the function should only be a function of the conformal time. That is, $f(\eta, x) = f(\eta)$. To fix this function as a function of the conformal time, we make use of our notion of distance. To ensure that the distance between neighbouring disentanglers on the same level is the same, we require that $f(\eta) (\Delta x)^2 = f(\eta') (\Delta x')^2$. If we set our conformal time to scale linearly with the difference between the comoving coordinates of neighbouring tensors, we get the condition $f(\eta) \sim \eta^{-2}$. Fixing the constant, we get the final form of the metric as :
\begin{equation}
    ds^2 = \frac{L^2}{\eta^2} \Big(- d\eta^2 + dx^2 \Big)
\end{equation}
Note that the form of the metric is the same as that of eq. (\ref{dS-metric}). Hence, the metric on the MERA geometry is given by that on a hyperbolic space -- that of the flat slicing of a de Sitter space. Our proposal will heavily rely on this result. 

\subsection{MERA for thermal states}
\label{thermalMERA}

Until now, we have been considering implementing the MERA on a pure state. Now, we shall see how to construct a MERA network on a thermal state. The essential difference is that when one considers the entanglement entropy of a subregion ($A$), one needs to consider the contribution arising from the fact that the state being considered is not pure : $S_A = S_\text{MERA} + S_\text{thermal}$. While implementing this in the original MERA framework, it was argued that after a finite set of steps (the number of steps goes as $\log{\beta}$), the coarse grained density matrix completely factorizes into maximally mixed states and one ends the MERA network geometry at the `wall' \footnote{As is noted in the discussion to follow, as of today we know that this scheme is valid only as an approximation. We thank Markus Hauru for pointing this out.}. In the context of the AdS/CFT correspondence, this relates to the bulk dual of a thermal CFT being a black hole \cite{Swingle:2009bg}. One now has to add an extensive component to the entropy calculation, arising from the fact that one started off with a thermal state. 

An alternative prescription was given in \cite{Molina-Vilaplana:2014mna}, wherein one deals with the case of a thermal state by linking the IR degrees of two MERA curtains. Let us now consider a generic mixed state given by a density matrix $\rho$ : we do that by enforcing the condition $\text{Tr}(\rho^2) < 1$. Now, we consider the purification $\ket{\Psi}$ of the density matrix, the state has to obey the condition $\text{Tr}_{\bar{A}}(\ket{\Psi}\bra{\Psi}) = \rho$. Such a state can be constructed by entangling every degree of freedom in the system with an accompanying purifying site. Hence, if we start off with a state with $N$ degrees of freedom, we end up with a purified state having $2N$ degrees of freedom. The purified state can be written as \begin{equation}
    \label{pure}
    \ket{\Psi} = \sum_{i_1 i_2 \ldots i_N \text{ ; } j_1 j_2 \ldots j_N=1}^{\chi}   c_{i_1 i_2 \ldots i_N \text{ ; } j_1 j_2 \ldots j_N } \ket{i_1 j_1} \otimes \ket{i_2 j_2} \otimes \ldots \ket{i_N j_N}
\end{equation}
It can be verified that this state can reproduce any density matrix for the region under consideration, by taking a trace over all the accompanying degrees of freedom. 
Given that we know that the MERA network geometry ends upon an output of a maximally mixed state, it would be beneficial to consider it. The density matrix for such a system with $N$ sites (in our case, at the curtain) is given by : 
\begin{equation}
   \rho_\text{wall} = \frac{1}{\chi^N} (\mathbb{I}_d)^{\otimes N}
\end{equation}
The entropy of such a state is given by $N\log{\chi}$. The purified state that one can write for such a density matrix is given by :
\begin{equation}
\label{MPDO}
    \ket{\Psi_\text{wall}} = \Big{(}\frac{1}{\sqrt{\chi}} \sum_{\alpha = 1}^{\chi} \ket{\alpha}_i \ket{\alpha}_j\Big{)}^{\otimes N}
\end{equation}

\begin{figure}[ht]

\centering
\includegraphics[width=0.5\textwidth]{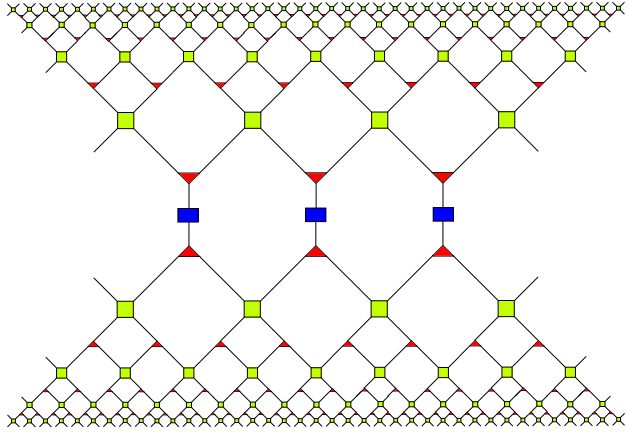}
\caption{The MERA for a thermal state factorizes into maximally mixed states after $\log{\beta}$ steps. Upon purification, corresponding IR sites are connected through each other with a MPS bridge (blue) which signifies entanglement between the IR sites.}
\label{figmixedMERA}
\end{figure}

The prescription suggested involves doubling the original MERA (signified by the purification), and `hanging' the accompanying MERA above the original MERA. With reference to eq. (\ref{MPDO}), we note the entanglement pattern in the IR sites of the two networks. To denote this in the network geometry, we link the corresponding IR sites of the two MERAs with an entangled bridge Matrix Product State (MPS) \footnote{Matrix Product States consist of a one dimensional array of tensors. Here, they carry a bond dimension of $\chi$.}. This procedure is illustrated in Figure (\ref{figmixedMERA}). 

While computing the entropy of a subsystem, the MPS links which are cut through by the minimal curve in the network act as a measure of the entropy arising due to the thermal nature of the system that one started off with. As mentioned earlier, this contribution can also be read off as the contribution by the horizon entropy of the AdS Black Hole. In our application of the hanging MERA prescription, we will not be motivated to adopt such a viewpoint.

\section{Towards a dS/MERA correspondence}
\label{sec4}
\begin{figure}[ht!]

\centering
\includegraphics[width=0.5\textwidth]{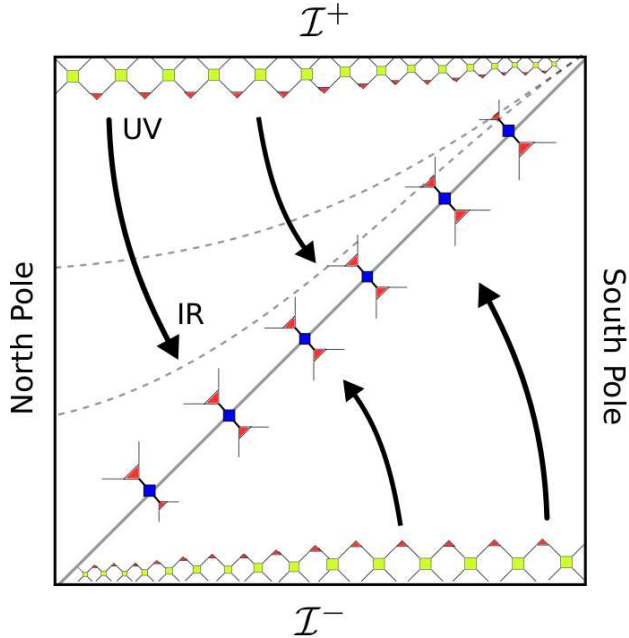}
\caption{The construction as proposed by our prescription for the dS/MERA correspondence. The UV degrees of freedom on both the conformal theory as well as its purification are seen lying on the conformal boundaries ($\mathcal{I}^\pm$). The renormalization flow in the network (as indicated by the arrows) evolves the network to the IR, wherein we obtain the factorized sites on both the sides. Here, the MPS states link corresponding sites to give us the `horizon'. This construction is overlaid on the flat slicing of the de Sitter Penrose diagram to demonstrate the correspondence. Note that the cosmological time increases while the conformal time decreases upon going from the IR to the UV.}
\label{construction}
\end{figure}

Given the success of the program that was kicked off by the AdS/MERA correspondence, a similar tensor network description for de Sitter spacetime would provide us with interesting insights on the emergence of de Sitter spacetime from entanglement as well as the hypothesized dS/CFT correspondence. Given the qualitative differences between the AdS and dS spacetimes discussed above, we anticipate such a construction to be more involved than the AdS construction. Given that the emergent dimension in de Sitter holography is the time dimension, the notion of causality will have to be faithfully recreated and reconstructed -- a task that was only recently undertaken for the AdS case \cite{May:2016dgv}. Below, we sketch one such possible route that can be taken towards working out a dS/MERA correspondence. The result stated below is illustrated in Figure (\ref{construction}).

\begin{itemize}
    \item Consider a conformally invariant quantum system in a thermal state \footnote{Such a system will generalize to a CFT in the thermal state, for the case of a continuum version of the MERA.}. We implement a MERA network on it, with a cap on the same (as described in Section \ref{thermalMERA}). As we have seen, the metric on such a tensor network, away from the cap will be the same as that of the flat FRW slicing of de Sitter.
    \item We shift coordinates on the MERA to the radial coordinates. Making use of the symmetry of the MERA we only keep the radial and time direction. 
    \item We change the holographic geometry of the MERA such that layers with equal coarse-graining (equal comoving radius) in the MERA lie along the equal time (equal comoving radius) curves as indicated in the upper triangle of the flat slicing of de Sitter in Figure (\ref{flatfigure}). We denote this system as $\mathcal{O}^+_\text{MERA}$.
    \item Since we started off with a thermal state, we now purify the state and produce a MERA for the accompanying sites as well. Carry out a similar modification in the holographic geometry as above. We denote this system as $\overline{\mathcal{O}^+_\text{MERA}}$.
    \item Now, we stitch up the IR of the $\mathcal{O}^+_\text{MERA}$ to the IR of the $\overline{\mathcal{O}^+_\text{MERA}}$ using MPS bridges, signifying entanglement between the corresponding sites in the IR.
\end{itemize}

If we are to associate the quantum system on which the MERA is in $\mathcal{O}^+_\text{MERA}$ with $\mathcal{I}^+$ in  and the quantum system on the accompanying purification sites with $\mathcal{I}^-$, it can be said that one recreates the Penrose diagram for a discretized version of the de Sitter spacetime. Such an identification is important for the correct running of the cosmological time. The metric and hence the causal structure is built in by construction, in the form of the tensor network that we have drawn. Note that the future event horizon is not covered in both $\mathcal{O}^+_\text{MERA}$ and $\overline{\mathcal{O}^+_\text{MERA}}$, since the MERA can't evolve a spacelike slice into a lightlike slice. However, the MPS states can be said to recreate the null surface thanks to a `gluing through entanglement operation' -- the entanglement between the two conformal systems being at play. 

Alternatively, one can carry out the same procedure as mentioned above but for the case of $\mathcal{O}^-_\text{MERA}$ (defined on $\mathcal{I}^-$) purified by $\overline{\mathcal{O}^-_\text{MERA}}$. It would be facilitated by making the coordinate transformations $t \rightarrow - t$ and $\eta \rightarrow - \eta$ so as to ensure that the time coordinate increases in the forward time direction. Analogously, we have the creation of the past event horizon through the MPS states. The choice between the two alternatives would depend on the intended use of the setup. The novel feature in our work is the fact that we obtain the entire Penrose diagram of a dS space from MERA.

Let us reiterate the difference between the standard AdS/MERA construction and the proposed prescription for the case of a dS/MERA correspondence given above. As noted in Sec.(\ref{sectionMERA}) for the more familiar case of AdS spacetime, the MERA prescription manages to recreate a (discretized) \emph{spatial slice} of the spacetime and not the entire spacetime -- only the radial direction being emergent. However, in  our proposal for obtaining a dS space from MERA we obtain the \emph{entire 'spacetime'}, with \emph{time} being the emergent direction. 

The proposed model could be said to be 'holographic' in the sense that given a state defined on the conformal theory living on $\mathcal{I}^+$ we can use the MERA prescription to obtain 'time evolution' of the discrete quantum fields in the bulk \footnote{If one was to write a Lagrangian for the conformal theory on the conformal boundary it would be 1+1 dimensional. However, we are not doing that in this paper. The 'time evolution' in the bulk spacetime obtained pertains to the entanglement in the state on the boundary as seen at different scales and \emph{not} a Lagrangian or Hamiltonian evolution of the boundary state itself.} Each constant time slice in the bulk would essentially contain information about the \emph{entanglement structure} in the boundary state at a corresponding scale. On a given time slice, one encounters an increasingly entangled state with the passage of cosmological time. As also seen in \cite{Beny:2011vh,Czech:2015kbp} the causal structure of MERA encourages us to consider dS spacetimes to be more natural representatives of the tensor network under consideration as compared to AdS spacetimes for which tensor networks depicting quantum error correcting codes \cite{Pastawski:2015qua,Almheiri:2014lwa} are deemed more fit.

Due to the nature of the construction, the identification of antipodal points on the two conformal boundaries checks out as suggested by \cite{Strominger:2001pn}. The notion of our cosmological time coincides with the notion that late times in the bulk correspond to the UV of the boundary theory, while early times correspond to the IR \cite{Strominger:2001gp}. While our construction was inspired by the hypothesized dS/CFT correspondence, we do note that our notion of holography obtained through the construction is different from the standard dS/CFT \cite{Strominger:2001pn, Maldacena:2001kr} correspondence given that we propose a mechanism for obtaining the bulk states motivated by the \emph{gravity from entanglement} \cite{VanRaamsdonk:2010pw} perspective as compared to primarily establishing a duality at the level of operators in conformal theory and the fields in the bulk dS spacetime. As is evident from our proposed construction, we see that the state of the conformal theory on the boundary that we start from is essentially described by a mixed state. As discussed in \ref{de Sitter}, this feature in our proposal follows the general structure suggested in \cite{Balasubramanian:2002zh} for the consistency of dS/CFT correspondence itself -- the dual theory to the bulk is given in terms of two disjoint but entangled conformal theories. The implication of this feature on the thermodynamics of the spacetime and cosmology would be discussed in future work. 

\section{Discussion}
\label{discussion}
Despite the highly suggestive nature of the construction proposed, the work done remains speculative in nature largely due to our lack of availability of strong checks in de Sitter holography \cite{Anninos:2012qw}. However, the novel feature of the work is the explicit involvement of the `gravity from entanglement' as well as a tensor network perspective for de Sitter facilitated by the hyperbolic form of the metric as defined on the MERA network geometry. 

An obvious direction of extension of the work would be to uplift the discrete MERA to its continuum version (cMERA) as done for the case of the AdS/MERA correspondence \cite{Haegeman:2011uy,Nozaki:2012zj}. However, such an exploration may face issues due to the presence of horizons and nature of the conformal boundaries. Similarly, it would be interesting to check the compatibility of our construction with the quantum error correction code perspective on holography \cite{Pastawski:2015qua,Verlinde:2016toy}. If our assumption of the conformal theories on the boundaries having a thermal density matrix holds, it brings about the question of how to interpret the temperature. Naively, it might be suggested that the temperature is related to the de Sitter radius (alternatively, the Hubble constant defined as : $H \sim L^{-1}$).  Since dimensionally we'd have $T \sim H$, a faster rate of expansion would terminate the network at a shallower scale. This would give rise to more sites in the IR which have a maximally mixed density matrix. We would like to interpret this as the observer along the North Pole losing the ability to influence more regions on $\mathcal{I}^+$, in the same time interval. 

It is to be noted that argument in  \cite{Swingle:2009bg} (and by extension, our construction) for the thermal MERA holds only as an approximation. As recently shown in the numerical scheme run in \cite{2015PhRvL.115t0401E}, the MPS tensors would need be connected to each other -- there is some entanglement between the degrees of freedom on the `horizon' \footnote{Technically, the resultant tensor network on the horizon would be formed out of Matrix Product Operators (MPO).}. Perhaps such a development would lead us to direct geometric realization of the event horizons in the tensor network, but due to the lack of a prescription to generate such a structure, we are unable to make progress on the same. 

The origins of the entropy of the cosmological horizon in de Sitter have proven to be pretty elusive, despite the fact that there have been many approaches which have been suggested \cite{Balasubramanian:2001rb}. Since the conventional AdS/CFT notions of holographic entanglement entropy don't seem to very straightforwardly generalize or hold relevance in the case of the entropy of the cosmological horizon as observed by an observer in the static patch of de Sitter \footnote{Though \cite{Sato:2015tta, Narayan:2015vda} talks of the Ryu-Takayangi proposal in the context of dS/CFT through analytic continuation of a certain higher spin theory in the EAdS/CFT case, it is not seen how it is of relevance to the cosmological horizon entropy unlike the case of the AdS Black Hole.}, it might be essential to rethink our notions of how we interpret this entropy in the terms of a bulk-boundary duality. We would like to note that the entropy as obtained in \cite{GibbonsHawking} is read off the area of the spacelike $codimension-2$ surface surrounding the observer, and this quantity remains the same throughout time for a given Hubble factor. From the tensor network perspective as well as the gravity from entanglement perspective, we would like to conjecture that the horizon entropy arises because of the the loss of influence on the boundary conformal theory with time. Since the evolution in time is intimately linked to the scale at which entanglement is removed by the MERA for our boundary theory -- earlier times corresponding to disentangling across a larger scale on the boundary theory, this would support the idea that the horizon entropy indeed arises from the entanglement across the horizon. By drawing a $codimension-2$ ball with $r=L$ around an observer, we could count the disentangler links cutting through the horizon. Whether such a counting would give us the correct scaling for the horizon entropy would be explored in the course of our forthcoming work.

We know that the early and late time geometry of the universe is described by approximately de Sitter phases \cite{Baumann:2009ds,Riess:1998cb}. Hence the correspondence discussed in this paper may help us to realize cosmological models in such a framework. Work done in \cite{Verlinde:2016toy} could mark the beginning of explorations in cosmology using tensor network holography. It would be interesting to work out how perturbation theory would work out on a discrete geometry constructed as per our proposal. Shifting from the comoving gauge to the spatially flat gauge can be thought of a local translation in time \cite{Cheung:2007st}. Since the emergent direction in our construction is time itself, we observe that the discrete local conformal transformations implemented in the MERA in \cite{Czech:2015xna} could eventually lead us to understand the gauge fixing in the context of `cosmological perturbation theory' on tensor network geometry. 

\textbf{\textit{Note}} After the completion of this work, there appeared an article \cite{Bao:2017iye} which tried to construct a quantum circuit toy model for expanding cosmologies following the \emph{gravity from entanglement} perspective. Although their model aims to make predictions for quasi-de Sitter spacetimes (which are well suited for inflationary cosmology), they seem to obtain the general features that we have obtained through our construction -- that of time evolution in the bulk being obtained by looking  at an increasingly entanglement state (as we move towards smaller length scales).

\acknowledgments

The authors would like to thank the anonymous referee for drawing our attention to some of the issues. We are also thankful to Lubo{\v s} Motl for his valuable comments on the initial idea. We also appreciate the correspondence and questions addressed by Markus Hauru and Javier Molina-Vilaplana, the encouragement provided for pursuing work in such a direction by Onkar Parrikar and Brian Swingle as well as the proofreading done by Kartik Sreedhar. RK thanks the hospitality of the String theory group at the Harish-Chandra Research Institute, Allahabad and the CFP, University of Porto where a part of this work was inspired.  


\end{document}